# Interactive molecular dynamics in virtual reality for accurate flexible protein-ligand docking

Helen M. Deeks[a,b,c], Rebecca K. Walters[a,b,c], Stephanie R. Hare[c], Michael B. O'Connor[a,b,c], Adrian J. Mulholland*[c] and David R. Glowacki*[a,b,c]

[a]Intangible Realities Laboratory, University of Bristol, School of Chemistry, Cantock's Close, Bristol BS8 1TS, UK; [b]Department of Computer Science, University of Bristol, Merchant Venturer's Building, Bristol BS8 1UB, UK; [c]Centre for Computational Chemistry, School of Chemistry, University of Bristol, Cantock's Close, Bristol BS8 1TS, UK

*adrian.mulholland@bristol.ac.uk, *glowacki@bristol.ac.uk

Simulating drug binding and unbinding is a challenge, as the rugged energy landscapes that separate bound and unbound states require extensive sampling that consumes significant computational resources. Here, we describe the use of interactive molecular dynamics in virtual reality (iMD-VR) as an accurate low-cost strategy for flexible protein-ligand docking. We outline an experimental protocol which enables expert iMD-VR users to guide ligands into and out of the binding pockets of trypsin, neuraminidase, and HIV-1 protease, and recreate their respective crystallographic protein-ligand binding poses within 5 – 10 minutes. Following a brief training phase, our studies shown that iMD-VR novices were able to generate unbinding and rebinding pathways on similar timescales as iMD-VR experts, with the majority able to recover binding poses within 2.15 Å RMSD of the crystallographic binding pose. These results indicate that iMD-VR affords sufficient control for users to carry out the detailed atomic manipulations required to dock flexible ligands into dynamic enzyme active sites and recover crystallographic poses, offering an interesting new approach for simulating drug docking and generating binding hypotheses.

## 1      Introduction

Computational researchers across a wide range of fields are becoming increasingly aware of their responsibility to explore low-cost simulation methodologies whose energy and hardware demands are environmentally sustainable. [1] Whilst the molecular dynamics (MD) approach to biomolecular simulation [2-5] and protein ligand-binding [6-12] has exploded in recent years, [13] the computational cost of MD-based approaches remains significant – a result of the fact that proteins are high-dimensional systems characterized by many local energy minima separated by a rugged landscape.

Building on previous work constructing interactive simulation frameworks (e.g., led by Brooks [14-16], Wilson [17], Schulten [18, 19], and others [20-22]), we have been exploring interactive molecular dynamics in virtual reality (iMD-VR) as a low-cost strategy for investigating biomolecular problems like protein-ligand binding. Part of the attraction of an MD-based approach is the fact that it can capture movement and flexibility of both the protein and the ligand. Our open-source iMD-VR framework Narupa [23] allows users to interactively visualise and manipulate the molecular dynamics of real-time simulations within a virtual environment with atomic-level precision. The iMD-VR approach recognizes that the neural nets of the human brain, trained over several aeons, offer extremely sophisticated machinery for efficiently undertaking tasks linked to 3D insight, spatial navigation, and manipulation, with energy demands that are a fraction of those required by sophisticated hardware-accelerated machine-learning frameworks. Narupa enables humans to utilize their 3D spatial awareness skills, and their ability to undertake 'on-the-fly' reasoning, to perform sophisticated operations on complex three-dimensional molecular structures, giving them the ability to move molecular systems between different regions of configuration space. In previous work, we have demonstrated that iMD-VR has acceleration benefits for various 3D molecular tasks compared to two-dimensional interfaces. [23, 24] The ability to physically reach out and manipulate simulated systems as if they were tangible objects gives users the opportunity to explore molecular transformations, mechanisms, and rare events that may otherwise be inaccessible using conventional high-performance computing (HPC) MD simulations. While we have previously demonstrated the potential of iMD-VR to generate dynamical pathways in small systems, it remains to be seen whether iMD-VR is sufficiently intuitive and enables enough control for researchers to undertake more complex tasks.

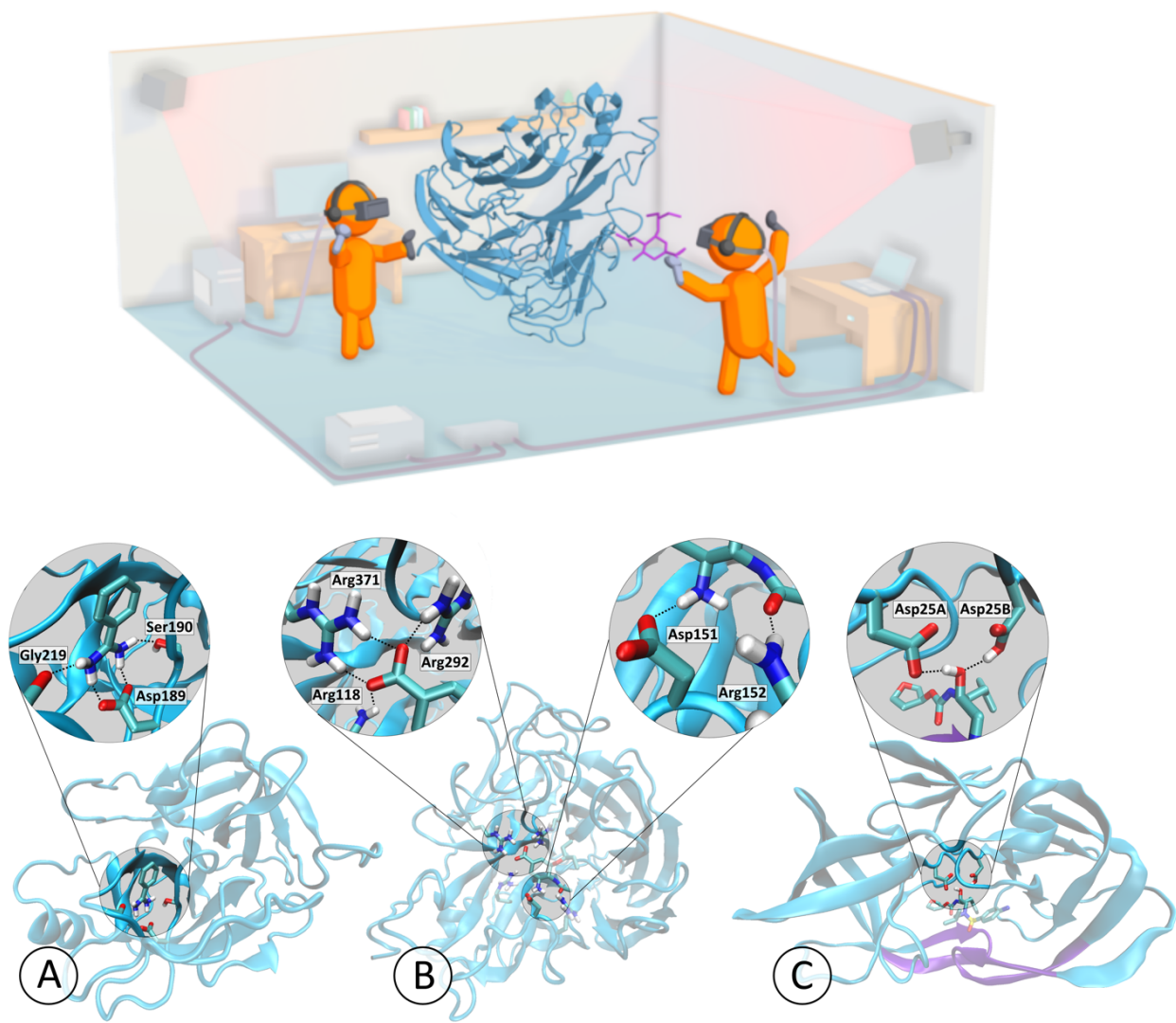

**Figure 1: Interactive protein-ligand rebinding using iMD-VR.** The top panel shows a schematic of Narupa, the open-source multiperson iMD-VR framework used to carry out the studies described herein, showing two participants using handheld wireless controllers to manipulate a real-time MD simulation of neuraminidase and oseltamivir. More information on the Narupa setup is available in ref 15. The bottom panel shows a three-dimensional representation of the binding pockets of each of the three protein systems interactively undocked and redocked, each bound to a ligand. The trypsin graphic (A) highlights the network of hydrogen bond interactions from Asp189, Ser190, and Gly219 to benzamidine (based off PDB code 1S0R). The neuraminidase graphic (B) highlights the hydrogen bonds between oseltamivir and both the 150-loop and a positively-charged trio of arginine residues of neuraminidase (based off a mutated neuraminidase structure derived from PDB code 2QWK). The HIV-1 protease graphic (C) highlights the hydrogen bonds between the hydroxyl group of amprenavir with both Asp25A and the protonated form of Asp25B (based off PDB code 1HPV). For HIV-1 protease, the protein 'flaps' have been highlighted in purple.

In this article, we focus on applying Narupa to unbinding and rebinding small molecules to proteins, as illustrated in Figure 1. Applying iMD-VR to protein-ligand systems requires moving a simulation between thermodynamically and kinetically distinct states (i.e., bound and unbound). This requires sophisticated three-dimensional spatial reasoning: the protein binding pocket needs to be located, and the ligand orientated such that key contacts to specific residues are properly recreated. We evaluate the utility of iMD-VR to facilitate such manipulations through applications to three protein-ligand systems of increasing complexity, and investigate the extent to which iMD-VR 'experts' and 'novices' were able to reversibly generate protein unbinding and rebinding pathways and recover crystallographic ligand poses for the systems shown in Figure 1. The results of the experimental protocol outlined herein show that iMD-VR users can quickly unbind and rebind ligands to proteins and recover experimental protein-ligand structures. For example, we show that an hour-long training session with an expert user enables novice iMD-VR users to recover the original binding poses for several different protein-ligand systems. We have established that the pathways generated by iMD-VR users are reproducible, and that the binding poses they recover are stable when used to initiate MD simulations over longer simulation timescales.

iMD-VR represents a relatively new approach to simulating drug binding, since it involves a fully (or predominantly) flexible dynamic system and also does not require predefining a CV along which to carry out biased molecular dynamics simulations. [25] Instead, the researcher simply 'draws' the desired pathway by manipulating a complex system between different states. This approach has the potential to accelerate the exploration of pathways through high-dimensional configuration space. Overall, our tests show that a well-designed iMD-VR setup enables expert and novice users alike to accelerate complex 3D state changes of biomolecular systems. Our results suggest that iMD-VR tools are sufficiently intuitive and provide adequate control to enable users to recover experimentally derived bound complexes. Moreover, the fact that the sampled pathways are reproducible suggests that they are not too far from the equilibrium ensemble.

## 2  Methods

**2.1 System Selection**

The systems selected for the studies outlined herein include, at one extreme, those that have well-characterized binding modes, where a user with prior knowledge of the system can use iMD-VR to sample binding and unbinding pathways. At the other extreme, we sought to push the limits of the iMD-VR tools and examine more complex protein-ligand systems. The three systems we chose are shown in Figures 1 and 2 and are described below in order of increasing complexity.

*2.1.1 Trypsin*

The first protein-ligand system is the enzyme trypsin with a benzamidine ligand. Trypsin is a well-studied serine protease hallmarked by a catalytic triad consisting of histidine, serine and aspartic acid. As shown in Figure 1A, a secondary motif of the trypsin binding pocket is an additional aspartic acid residue (at position 189) that stabilises the positively charged amino acids, binding a peptide chain in a pose that allows for specific cleavage of the carboxyl-side chain in which these amino acids are present.[26] Asp189 is charged, partially-buried, and is surrounded by a water network that favours desolvation.[27-29] Trypsin inhibitors can exploit this by mimicking positively-charged residues and forming an electrostatic contact with Asp189, an interaction further stabilized by a network of hydrogen bonds to Ser190 and Gly216 (Figure 1A).[30] The binding mode between trypsin and one of its inhibitors, benzamidine, has relatively fast kinetics and is therefore often applied as an initial use case for theoretical methods.[31-34] From an iMD-VR perspective, the trypsin-benzamidine system represents a binding mode where users do not need to induce significant conformational changes in the protein, and the ligand has only one rotatable bond. In effect, a user need only move the benzamidine out of the binding pocket before placing it back, making trypsin a relatively simple test case for iMD-VR.

Figure 1A shows a three-dimensional binding mode of benzamidine in the secondary binding pocket of trypsin, illustrating how residues Asp189, Ser190, and Gly216 bind to ligands (based off PDB code 1S0R). Figure 2A shows the two-dimensional binding mode of two ligands, benzamidine and indole-amidine, into trypsin. Supplementary video A.1 (https://vimeo.com/354833443) shows an example of an expert iMD-VR user unbinding and rebinding benzamidine from trypsin in iMD-VR.

*2.1.2 Neuraminidase*

H7N9 neuraminidase is a glycoside enzyme found on the surface of influenza virions that catalyzes the hydrolysis of sialic acid residues, a process integral to influenza virus mutation.[35, 36] This strain was found in the influenza virus responsible for the 2013 bird flu pandemic.[37] The process of influenza virus replication begins with the virus attaching itself to the cell via the binding of hemagglutinin (a viral cell surface protein) to sialic acid, found on the end of glycoproteins attached to the cell membrane. Once anchored, the virus can enter the cell and hijack its machinery in order to replicate itself. After new viruses have been manufactured, neuraminidase proteins cleave sialic acid groups from cellular glycoproteins, breaking the anchor between the virus and the host cell, freeing the new viruses to infect other cells. In recent years, there has been extensive research on the discovery of inhibitors that mimic sialic acid, thereby preventing the cleavage of sialic acid residues by neuraminidase and obstructing the release of new viruses. [38, 39] Oseltamivir (Tamiflu) is a transition state analogue of sialic acid and licensed therapy for influenza that binds to a triad of arginine residues in the neuraminidase active site (Arg118, Arg292, and Arg371) via its carboxylate group. [40] In addition to this, a crucial loop opening and closing mechanism, known as the 150-loop motion, is involved in the unbinding and rebinding of oseltamivir (Figure 1B). [41, 42] From an iMD-VR perspective, the binding of oseltamivir to neuraminidase represents a more complex task than the trypsin-benzamidine system for two reasons: (1) Given that 150-loop movement is imperative to binding, this backbone motion needs to be carried out by the iMD-VR user; (2) With eight rotatable bonds, oseltamivir has considerably more conformational flexibility than benzamidine, making it more of a challenge to re-establish the correct bound configuration.

Figure 1B shows a three-dimensional binding mode of oseltamivir in the active site of neuraminidase, illustrating how the 150-loop and arginine trio binds to ligands (based off a mutated neuraminidase structure derived from PDB code 2QWK; details for how this was done can be found in ref [37]). Figure 2B shows the two-dimensional binding mode of two ligands, oseltamivir and zanamivir, into neuraminidase. Supplementary video B shows an example of an expert iMD-VR user unbinding and rebinding oseltamivir from neuraminidase in Narupa.

*2.1.3 HIV-1 protease*

HIV-1 protease is a viral aspartyl protease essential to the life cycle of HIV, responsible for cleaving precursor polypeptides into functional proteins, making it an attractive drug target for preventing HIV maturation. Structurally, HIV-1 protease is a homodimer that shares a single active site between two protein subunits, each of which contributes a catalytic aspartic acid. Mechanisms for HIV-1 protease cleavage of precursor proteins have been proposed [43]; however, it is broadly understood that the two aspartic acid residues each act as an acid or base respectively, activating a water molecule that then goes on to break peptide carbonyl bonds via nucleophilic attack. Sulfonamides are a class of drug licensed for the treatment of HIV that hydrogen bond to the catalytic aspartic residues and thus block protease activity [44], an example of which is amprenavir. Functionally, the two aspartic acids are thought to primarily exist in a monoprotonated state, especially when in the presence of an inhibitor [45], although it has been debated which tautomer favours amprenavir binding. [46-50] The HIV-1 protease active site is gated by two beta-hairpin flaps that shift through a series of different conformational states before ligand binding. [51] Therefore, this task requires iMD-VR users to move these flaps into an open position to guide amprenavir out, and then carefully place the loops back without disrupting their secondary structure. From an iMD-VR perspective, the motion of these flaps, combined with the higher rotational flexibility of amprenavir, makes this a particularly challenging binding task. Figure 1C shows a three-dimensional binding mode of amprenavir in the active site of HIV-1 protease, illustrating how Asp25A and Asp25B bind amprenavir (based off PDB code 1HPV). The active site flaps are coloured in purple. Figure 2C shows the two-dimensional binding mode of amprenavir into HIV-1 protease. Supplementary video C (https://vimeo.com/354834013) shows an example of an expert iMD-VR user unbinding and rebinding amprenavir from HIV-1 protease in Narupa.

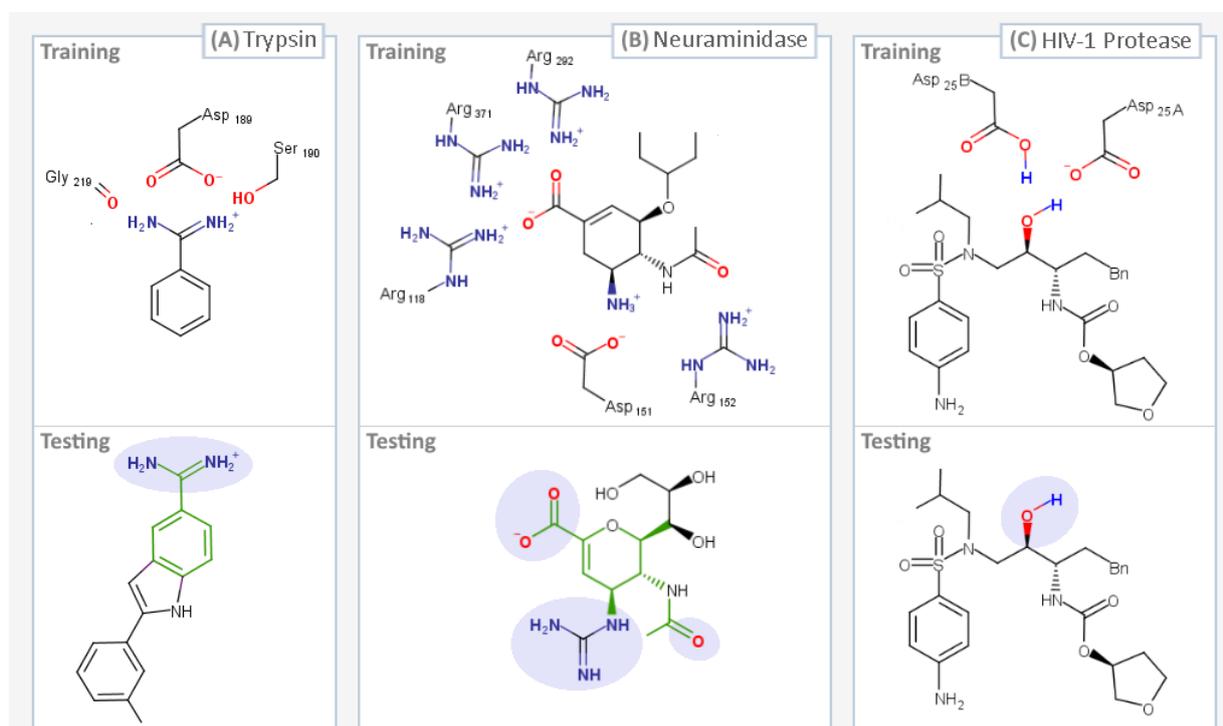

Figure 2: **Summary of iMD-VR docking tasks.** The ligands that were interactively unbound and rebound in a series of user tests, two for each of the three systems: Trypsin (A), Neuraminidase (B), and HIV-1 Protease (C). For each protein system, we devised two docking tasks, a training phase (where a trace representation of the ligand in the bound pose was present to guide users as shown in supplementary videos A - C), and a testing phase, where no trace atoms were shown. For the testing tasks, parts of the ligand which interact with key residues have been highlighted in lilac. For proteins A and B, where the testing task ligand differs from training, the shared scaffold between the two ligands is highlighted in green

## 2.2 Overview of Binding and Rebinding Tasks

Using these candidate protein-ligand systems, we carried out iMD-VR tests. Users were split into expert and novice cohorts. Expert users had extensive experience using iMD-VR to interact with molecules and had some degree of familiarity with the protein-ligand systems described in this paper. The experts first tested the viability of using iMD-VR to generate ligand unbinding and rebinding pathways. Expert iMD-VR studies were carried out to establish whether the original crystallographic structure could be re-established for each task. Then, a cohort of novice users were asked to carry out a series of unbinding and rebinding tasks, in order to control for the experts' familiarity with iMD-VR tools and the selected systems. This novice users had little experience using iMD-VR but did have a scientific background in biomolecular simulations.

### 2.2.1 Expert user tests

Expert iMD-VR users were tasked with unbinding and rebinding three ligands from the binding pockets of three proteins: trypsin, neuraminidase, and HIV-1 protease, detailed in Section 2.1. Figure 1 shows a three-dimensional representation of each of the three experimentally derived poses that expert users were asked to recreate. In these tasks, experts were initially given a faint trace of the ligand atoms in the correct pose, providing a visual indication of where the ligand sits in the crystallographic binding pose. Experts could freely switch between interacting with either a single atom at a time or a group of atoms together (e.g., moving one atom of the ligand or the center of mass of the entire ligand at once). Experts could also add or remove positional restraints to atoms, such as when manipulating the active site flaps during the HIV-1 protease task. Supplementary videos A-C show experts unbinding and rebinding small molecules from trypsin, neuraminidase and HIV-1 protease, and also show the guide trace for the position of the ligand atoms. [23]

### 2.2.2 Novice user tests

Once the iMD-VR unbinding and rebinding pathways in Narupa had been established by the expert users, novice users were recruited to complete unbinding and rebinding tasks with each of the three protein systems. Novice users interacted with each system until they felt they had correctly established a binding pose, spending approximately five minutes on each task. The protein-ligand systems were given in order of increasingly complexity. Figure 2 gives a summary of tasks novice participants were asked to complete. We recruited a total of ten novice participants in two cohorts of five ('cohort one' and 'cohort two'). Cohort one completed tasks described in 2.3.1 – 2.3.4. Cohort two completed tasks described in 2.3.1 – 2.3.4 and 2.3.6. More detailed demographic and task information can be found in the ESI.

To ensure there was a comparable level of proficiency with the Narupa iMD-VR interface prior to attempting protein-ligand binding, novice participants were first placed in a simulation of $C_{60}$ molecules and given an opportunity to familiarize themselves with the controls and the interaction, as was done in previous studies. [24] Additionally, before interacting with each of the three protein systems described in Section 2.1, we took advantage of the multiplayer feature of Narupa: novice users were placed in the same iMD-VR simulation as the expert user (as shown in the Figure 1 schematic), where they could see a fully three-dimensional representation of the bound poses shown in Figure 1. To ensure the novices had a comparable level of knowledge of each system, while co-habiting the simulation space, the iMD-VR expert gave a brief background on the system and highlighted the interactions shown in Figure 1.

Once the system had been introduced, participants were asked to complete a 'training' ligand unbinding-rebinding task. During the training phase, novices had a trace of the ligand atoms in the bound pose to help guide them. Next, participants were given a 'testing' ligand unbinding and rebinding task. A summary of the training and testing tasks for each of the three protein systems is shown in Figure 2. In these experiments, no trace atoms of the ligand were present, meaning that participants had to rely more on their chemical insight to find the correct binding pose. To control for learning effects during the testing phase of trypsin and neuraminidase binding, novices were asked to bind different ligands from those used during training. The alternative ligands were chosen on the basis of key interactions remaining between the ligand and protein in the training portion of the study (highlighted in lilac in Figures 2A and 2B) and sharing structural similarities (highlighted in green in Figures 2A and 2B). Due to a lack of availability of an appropriate alternative ligand to amprenavir, and because of the heightened complexity of the task owing to the number of rotatable bonds amprenavir contains, amprenavir was used for both the training and testing phases of the HIV-1 protease system. Novice users also had a more limited user interface, where they only had the option of interacting with a single atom at a time and could not apply or remove positional restraints.

## 2.3 iMD-VR simulation set up

The Narupa selection utility, shown in supplementary videos A – C, allows users to select and apply visualization and interaction settings to a subset of atoms in a system. For all systems, the following selections were created: (a) protein backbone atoms, (b) key active site residues, (c) ligand atoms. Each system had some specific features, detailed below. Further simulation-specific details can be found in the ESI.

*2.3.1 Trypsin and benzamidine (expert task and novice training task)*

This task (Figure 2A, 'Training') was completed by both experts and novices. Given the relatively simple binding mode between benzamidine and trypsin, where the protein does not need to be manipulated by the user, the entire protein backbone was held in place with a positional restraint (details can be found in the ESI). To enable the user to visually identify the location of the active site in the protein, residues Asp189 and Ser190 were fully rendered and the backbone oxygen of Gly219 was shown, similar to the representation shown in Figure 1A. Additionally, supplementary video A.1 shows the rendering scheme used for trypsin and benzamidine. For the remaining residues, only the protein backbone was rendered. A trace of benzamidine in the crystallographic pose was used as a visual indication of where the ligand binds, encouraging the user to re-establish the original bound pose as closely as possible.

*2.3.2 Trypsin and indole-amidine (novice testing task)*

Following the training phase, novices were asked to bind indole-amidine into trypsin (Figure 2A 'Testing'). Indole-amidine was selected as an alternate ligand for novices to bind as it contains the benzamidine moiety and adopts the same binding mode to Asp189, Ser190 and Gly219. However, unlike benzamidine, indole-amidine has a non-symmetrical extended scaffold that needs to be correctly orientated. This testing task used the same protein restraints and rendering scheme as the training task with trypsin and benzamidine (Section 2.3.1). However, no trace atoms were shown.

*2.3.3 Neuraminidase and oseltamivir (expert task and novice training task)*

This task was completed by both experts and novices. Binding oseltamivir requires a small amount of manipulation of the protein, as the positions of the 150-loop residues over the ligand need to be carefully maintained. However, as no significant shift in the tertiary structure of neuraminidase is required, all backbone atoms were held with a positional restraint (details can be found in the ESI). To aid in ensuring the 150-loop was correctly placed, two strategies were adopted. First, residues Asp151 and Arg152 of the 150-loop were fully rendered so the user could see their position. Second, a trace representation of the position of the 150-loop residues in the closed position was rendered so the user could see where they should be placed. Additionally, to show if a hydrogen bond between the carboxylic acid of oseltamivir and the arginine trio of the neuraminidase binding pocket had been established, Arg118, Arg292, and Arg371 were also fully rendered. For the remaining residues, only the backbone atoms were shown. A trace of the oseltamivir atoms in the correct position was used as a visual indication of where the ligand binds, ensuring that the original binding pose could be closely replicated. Supplementary video B shows the rendering scheme used in Narupa for neuraminidase and oseltamivir.

*2.3.4 Neuraminidase and zanamivir (novice testing task)*

Following the training phase, novice users were asked to bind zanamivir to neuraminidase. Zanamavir was selected as the second binding task as it is a chemical analogue to oseltamivir, sharing the same ring scaffold, carboxylic acid group, and hydrogen-bonding contacts to the 150-loop. However, zanamivir does exhibit some structural differences from oseltamivir; therefore, this task requires the user to recognize the similarity in binding modes between the two drugs. As discussed below, we used this particular task to examine the results obtained if we removed backbone restraints entirely. The same protein rendering scheme as neuraminidase and oseltamivir was used (Section 2.3.3), however, no trace atoms were present.

*2.3.5 HIV-1 protease and amprenavir (expert task)*

Binding small molecules to HIV-1 protease requires a significant shift in the backbone atom positions, altering the structure from closed to open. During the simulation, the user could toggle a positional restraint force on the active site flaps, allowing them to be held in place once an open or closed conformation had been established. An additional strategy to aid in the opening and closing of HIV-1 protease was to render trace atoms of the protein backbone in the position of the active site flaps in both the closed position and the open position, giving the expert user a visual indication of the range of motion the loop has between the two states, as well as allowing the user to closely re-establish the original closed conformation once amprenavir had been redocked; details on how this was done can be found in the ESI. Aside from the active site beta-hairpin flaps, which were manipulated by the expert user during the simulation, the protein backbone atoms were held by backbone restraints. Details on the positional restraints can be found in the ESI. Key contact residues Asp25A and Asp25B were fully rendered; the remaining amino acids only had their backbone atoms shown. A trace atom representation of amprenavir was used as a visual indication of where the ligand binds, ensuring the original bound conformation could be re-established.

*2.3.6 HIV-1 protease and amprenavir (novice testing and training tasks; cohort two only)*

Novice users from cohort two, totalling five participants, were asked to unbind and rebind amprenavir twice, once with the aid of trace atoms and once without. All other backbone atoms in the protein were positionally restrained. To streamline the task, novices were not required to open and close the HIV-1 protease flaps prior to and after binding amprenavir, instead starting from an HIV-1 protease conformation that had been pre-opened by an expert using iMD-VR and held with a positional restraint. Details of this can be found in the ESI. Both the testing and training binding tasks used the same rendering scheme as the expert user tests (Section 2.3.5) – the only difference between the two tasks was the presence of trace atoms.

*2.3.7 Backbone restraints*

During iMD-VR, users can apply force to any atom in the system – including those integral to the protein tertiary or quaternary structure. In order to ensure that the protein tertiary structure required for binding was not distorted, a positional restraint was applied to the backbone atoms in the proteins. To evaluate the impact of these restraints, we removed the backbone restraints for cohort two during the neuraminidase and zanamivir testing phase. A comparison of task completion between cohort one and cohort two for undocking and redocking zanamivir from neuraminidase can be found in Figure S3 of the ESI. These results, discussed in further detail in the ESI, show that careful iMD-VR users (expert and novice alike) were able to carry out binding and unbinding tasks without positional restraints. For analysis of iMD-VR trajectories, we observed that PCA carried out on iMD-VR trajectories which utilized positional restraints produced better results for the binding/unbinding pathways of interest, owing to diminished noise from the backbone fluctuations.

**2.4 Analysis of iMD-VR results**

For both the expert and novice interactive unbinding and rebinding tasks, trajectories of the protein-ligand system were recorded, taking a snapshot of the system every 0.25 ps. iMD-VR trajectories were analysed using an RMSD calculation protocol (described in the ESI). The RMSD was used to: (1) analyse the expert trajectories to see how the system is changed from a bound to unbound and back to bound state; (2) analyse novice trajectories to identify how close a user got to recreating the starting pose; and (3) analyse whether the recovered binding poses were stable. Specifically, we selected an iMD-VR frame with an RMSD close to the starting coordinates and used it to initialize a longer timescale (200 nanoseconds) MD trajectory. Further details can be found in the ESI.

To qualitatively assess the reproducibility of the binding pathways users explored using iMD-VR, we utilized a Principal Component Analysis (PCA) tool called *PathReducer*, [52] which takes as input an xyz file containing a series of molecular structures (in this case, an iMD-VR-generated trajectory) and outputs the set of principal coordinates which captures the most structural variance in the fewest coordinates. In effect, PathReducer collapses a 3N-dimensional (where N = number of atoms) trajectory into a visualisable path which spans two or three dimensions. Further details of this can be found in the ESI.

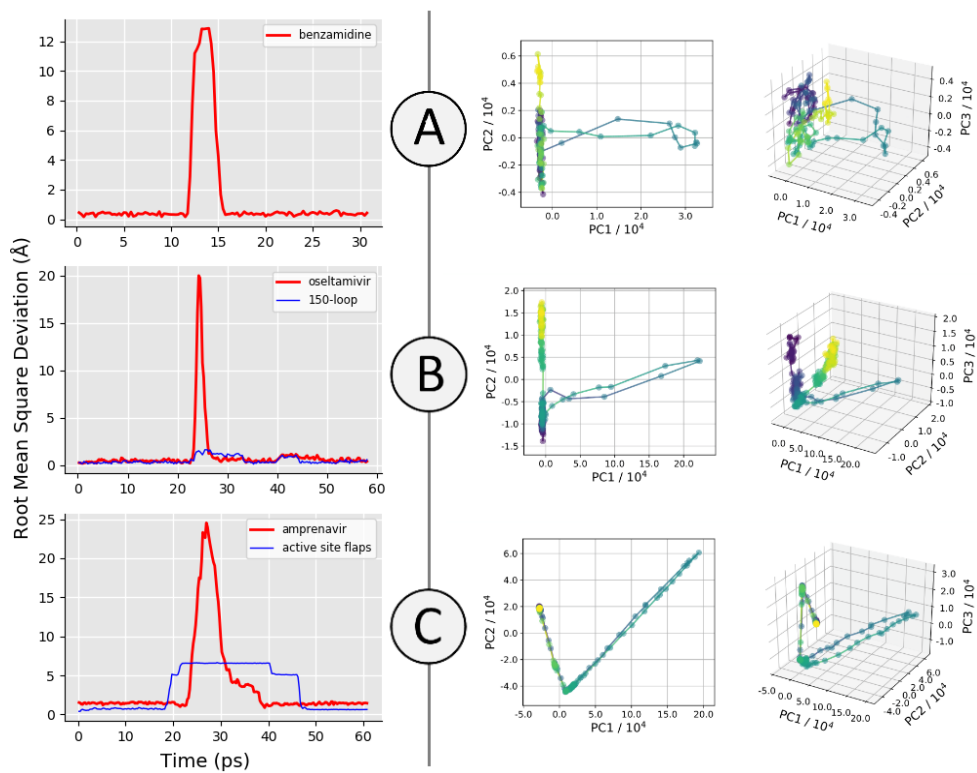

Figure 3: **Recovering binding poses of three protein-ligand systems using iMD-VR.** Left: Ligand RMSD of iMD-VR generated trajectories by experts (compared to starting coordinates): (A) trypsin and benzamidine, (B) neuraminidase and oseltamivir, and (C) HIV-1 protease and amprenavir. For neuraminidase, the RMSD of the 150-loop residues is shown in blue. For HIV-1 protease, the RMSD of the active site flaps backbone atoms is shown in blue. Right: The top two (left plots) and top three (right plots) principal components (PC1-3) for representative trajectories of the three protein-ligand systems. Time is represented for each trajectory by the colour of the plot, starting at purple for the beginning of the trajectory and yellow for the end, passing through blue and green in between.

## 3. Results and Discussion

### 3.1 Expert unbinding and rebinding tasks

*3.1.1 Trypsin*

Figure 3A shows the RMSD of the expert user-generated trajectory of benzamidine unbinding and rebinding to trypsin, alongside the top two and top three PCs along the trajectory path. The RMSD plot of benzamidine shows that the original binding pose was recovered by the expert user, as the RMSD between the ligand at the end of the trajectory and the beginning is very low. The PCA results for this system showed that the direction of PC1 (the PC that captures the most structural variance along the trajectory) corresponds predominantly to motion of the ligand away from the protein. The value of PC1 at the end of the trajectory is very similar to that at the beginning, which is more evidence that this user was able to replicate the original bound pose of the drug. PC2 and PC3 correspond to fluctuations of the backbone and side chains not in the active site of the protein, hence the trajectory not returning precisely to its initial value of these PCs. Supplementary animation A (https://vimeo.com/354828618) shows the trajectory generated by an expert user (31 picoseconds of simulation in total). Figure 4A shows the RMSD of an expert-generated rebound trypsin-benzamidine complex over 200 nanoseconds of production MD, after solvation, minimization and equilibration as described in the ESI, establishing that the iMD-VR produces stable poses for this system.

Expert study subjects were able to unbind benzamidine in under 5 picoseconds and obtain the rebound pose within another 5 picoseconds (Figure 3A). Using our computing architecture, as described in the ESI, we were able to achieve MD simulation rates of 4.45 picoseconds per minute of real time, meaning that benzamidine could be unbound and rebound on the scale of minutes. As such, this demonstrates the efficiency with simulations can be interactively pushed between two distinct states. Trypsin-benzamidine binding events have previously been observed to take nanoseconds of simulation time [32], and theoretical calculations of trypsin-benzamidine unbinding rate constants, $k_{off}$, predict millisecond dissociation times [31, 33]. To unbind or rebind benzamidine, the user does not need to make any major conformational changes to the binding site of trypsin, so completing this task was a matter of applying sufficient force to break the relatively weak electrostatic contacts between benzamidine and Asp189 and continuing to apply this force until the ligand had fully dissociated from the protein (supplementary video A.1). On rebinding, the user had to take care to keep the orientation of benzamidine such that the charged amidine group is pointing towards the bottom of the pocket. If the user attempted to bind benzamidine with the benzyl group directed towards the positively-charged Asp189 residue the ligand would be repelled out of the binding pocket, demonstrating the real-time responsiveness of iMD-VR to energetically unfavourable input, as shown in Supplementary video A.2. The video shows how, after being guided out of the binding pocket and replaced in an incorrect orientation, the N1 and N2 of benzamidine fail to form electrostatic contacts to Asp189, and is subsequently repelled out of the binding pocket, moving towards the HE1 of His57 in Trypsin, where it remains in a transiently stable pose over at least 200 ps, as shown in Figure S2-A1. Supplementary video A.2 shows the expert utilizing the 3D interface of Narupa to identify those amino acids with which the amidine group interacts with. This alternate binding mode, in which benzamidine favours the catalytic triad and polar residues surrounding the binding pocket is similar to previous observations by Buch et al. [32]. Figure S2-A2 shows how it was possible for the expert user to apply continuous force to the benzamidine when rebinding, guiding it past the cluster of hydrophilic groups and into the protein core and moving it out of the protein via an alternate opening. Whilst the kinetic barriers for these pathways are likely to be very large; they nevertheless show the ability of iMD-VR to quickly enable exploration of new dynamical pathways.

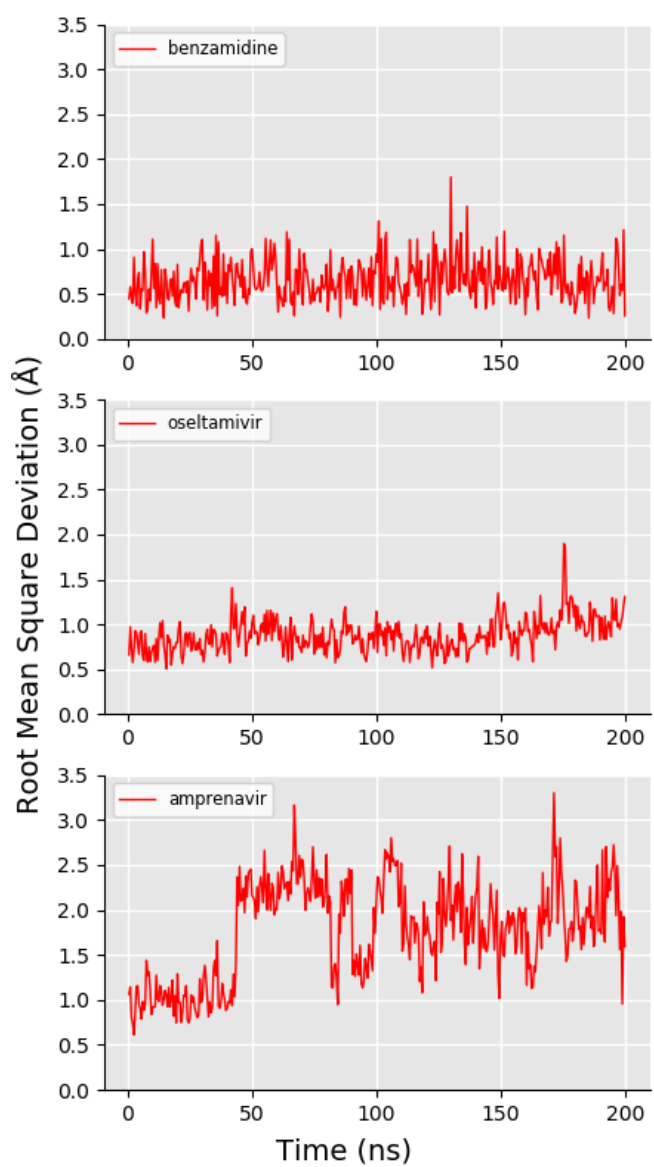

Figure 4: **Ligand RMSD over 200 nanoseconds of molecular dynamics.** Showing the RMSD of each of the three ligands (compared to the initial solvated, minimized and equilibrated bound coordinates) during 200 nanoseconds of molecular dynamics simulation. From top to bottom: (A) benzamidine (in trypsin), (B) oseltamivir (in neuraminidase), and (C) amprenavir in HIV-1 protease.

*3.1.2 Neuraminidase*

Figure 3B shows the RMSD of the expert-generated trajectory, alongside PathReducer-generated principal component plots, establishing that the original binding pose was recovered by the expert user. Similar to the trypsin system, PC1 is dominated by distances between atoms in the ligand and atoms in the protein, and so the variation along PC1 that can be seen along the trajectory path can be directly attributed to the removal of oseltamivir from the active site. PC2 and PC3 are again predominantly defined by distances between atoms in residues of the protein far from the active site, and thus fluctuations of the associated side chains lead to an offset in the pathway's start and endpoints. Supplementary animation B (https://vimeo.com/354829098) shows the trajectory generated by an expert user (58 picoseconds of simulation time in total). Figure 4B shows the RMSD of an expert-generated rebound oseltamivir-neuraminidase complex over 200 nanoseconds of production MD, after solvation, minimization and equilibration as described in the ESI, establishing that the iMD-VR can produce stable poses for this system.

Using our computing architecture, as described in the ESI, we were able to achieve MD simulation rates of 4.51 picoseconds per minute of simulation time, meaning that oseltamivir could be unbound and rebound on the scale of minutes of real time. Unbinding oseltamivir requires applying enough force to break any charged contacts or hydrogen bonds formed between the triad of positively charged arginine residues in the active site and the negatively charged carboxylate group of oseltamivir. Upon unbinding, the interactions from the amine group and carbonyl group of the drug and the two residues which comprise the 150-loop (Asp151 and Arg152 – see Figure 1B) were also broken, causing the loop residues to dynamically open as the drug leaves the active site as shown in supplementary video B (https://vimeo.com/354833800). The backbone of the 150-loop did not dynamically open when the drug exited the active site as the backbone was held in place with restraints. When removing oseltamivir from the binding site, the molecule would occasionally flip itself over, losing the bound orientation. As such, when attempting to re-establish the correct binding pose, the user needed to be extremely careful to ensure that binding interactions between the drug and the 150-loop residues were successfully recreated, allowing the drug to flip itself back into the correct pose; if the user was too forceful, the drug would enter the active site in the incorrect orientation. A representation of oseltamivir 'flipping' as it is being removed from the neuraminidase binding pocket can be found in Figure S2-B1. We observed that when oseltamivir is back into the correct bound conformation, the two 150-loop residues will also move back into the original configuration without any input from the user, demonstrating how iMD-VR can dynamically reform energetically-favourable contacts.

### *3.1.3 HIV-1 Protease*

Figure 3C shows the RMSD of the expert-generated trajectory, alongside generated principal component plots, establishing that the original binding pose was recovered by the expert-user. The PCA results in this case are unique, as the user had to move two flaps capping the active site prior to removal of the bound amprenavir (discussed in more detail below, see Figure 5). Each approximately linear segment of the PC plots corresponds to a conformational change event: First, one flap is moved away from the active site. Then the second flap is moved away from the active site. Finally, the ligand is removed from the active site. This process is then reversed to rebind the ligand. The plots show strikingly similar pathways for unbinding and rebinding, likely as the principal coordinates are each capturing a significant movement of the protein backbone and thus ignoring smaller residue fluctuations; comparatively, PC2 and PC3 in the trypsin and neuraminidase systems do not have the same starting and end points as they are 'picking up' protein fluctuations along the trajectory. Intuitively, PC1 is dominated by changes in distances corresponding to movement of the ligand out of the protein, PC2 corresponds predominantly to distances between the first 'flap' and other parts of the protein, and PC3 corresponds predominantly to distances between the second 'flap' and other parts of the protein. Supplementary animation C (https://vimeo.com/354829412) shows the trajectory generated by an expert user (61 picoseconds simulation time in total). Figure 4A shows the RMSD of an expert-generated rebound HIV-1 protease-amprenavir complex over 200 nanoseconds of production MD, after solvation, minimization and equilibration as described in the ESI. Some fluctuations to the RMSD value are observable. Inspection of the 200 nanosecond trajectory shows that the hydroxyl group 'flips' its orientation relative to the two catalytic aspartic acid residues, resulting in amprenavir adopting a slightly different binding pose throughout the simulation, albeit still remaining in the binding pocket of HIV-1 protease. Figure 5 shows a step-by-step breakdown of the process of unbinding and rebinding amprenavir from HIV-1 protease, establishing that the iMD-VR can produce stable poses for this system. In the *apo* form, the active site flaps of HIV-1 protease are thought to dynamically switch between various open and closed conformations [51]. Upon binding, the once-open flaps have a stabilizing effect by closing over the ligand. [53, 54] Experimental estimates of $k_{off}$ for amprenavir binding to WT HIV-1 protease correspond to an average binding residence time on the scale of seconds [55, 56], making it a difficult transition to simulate without some form of biasing. Using our computing architecture, as described in the ESI, we were able to achieve MD simulation rates of 4.45 picoseconds per minute of simulation time, meaning that amprenavir could be unbound and rebound – including opening and closing the active site flaps - on the scale of minutes of real time.

Additionally, the Narupa selection language enabled the expert user to remove positional backbone restraints and move the flaps into an open position, before replacing the restraints and extracting amprenavir away from the binding pocket. To rebind amprenavir, the hydrogen bonding to Asp25A and Asp25B was reformed and, once the stable binding configuration had been found, the user moved the beta-hairpin flaps back into the closed conformation, again replacing the restraints. We note that we initially performed this task with a tautomer of HIV-1 protease where neither Asp25 residues were protonated; in this state, amprenavir seemed to favour a slightly different binding pose as its hydroxyl group no longer acts as a hydrogen bond acceptor. To further investigate the difference in binding mode between the two tautomers of HIV-1 protease, both were placed in Narupa with all positional backbone restraints removed. The expert user observed both binding poses without interacting with the system over 125 picoseconds of simulation time. In the monoprotonated HIV-1 protease system, the hydroxyl group of amprenavir rigidly points towards the deprotonated aspartic acid residue. Conversely, when both catalytic residues are deprotonated, the hydroxyl group frequently flips between pointing at either aspartic acid, causing amprenavir to 'sag' in the active site (see Figure S2-C1). The positively charged hydrogen on the protonated aspartic acid appears to have a stabilising effect on the orientation of the hydroxyl group, whereas removing the hydrogen causes the hydroxyl group to more easily flip as one orientation is no longer electrostatically 'penalized' by the protonated residue. Nonetheless, it is worth highlighting that Narupa enables qualitative observation of such binding dynamics in real time, demonstrating that iMD-VR enables both manipulation of molecular dynamics and easy comprehension of distinct dynamical behaviour in systems that have only small structural differences.

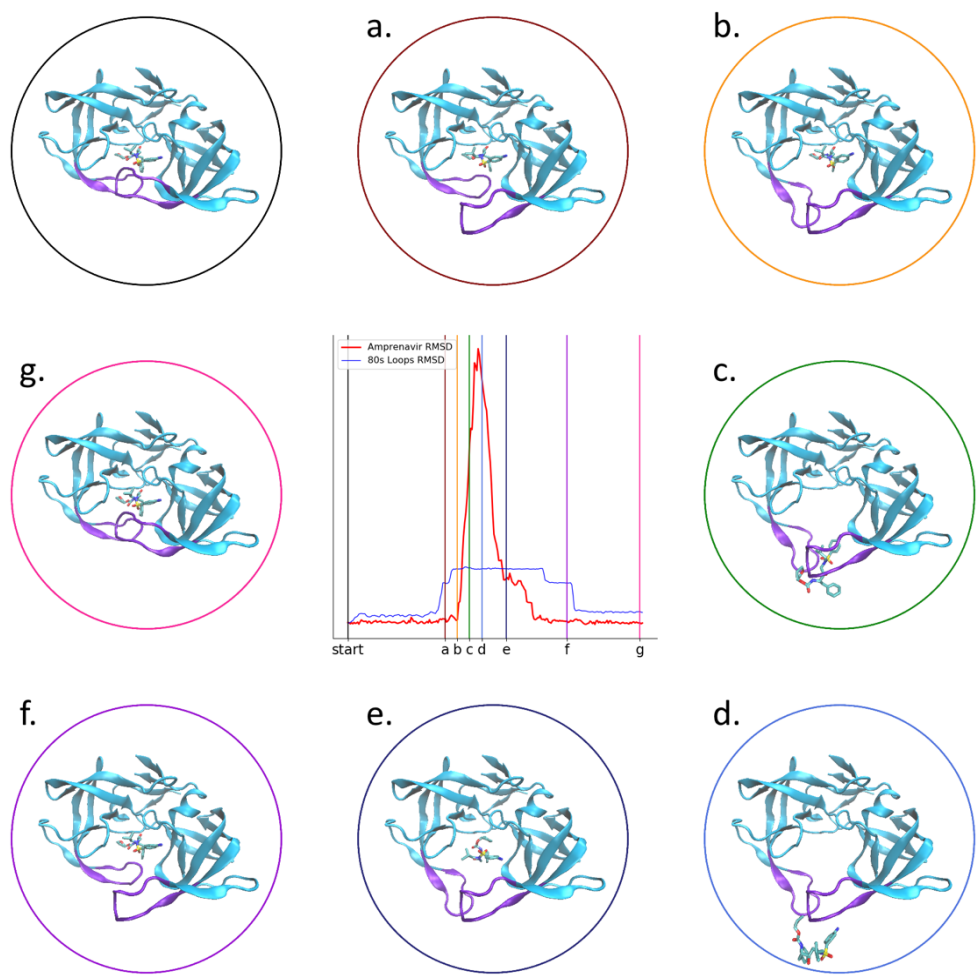

Figure 5: snapshots along the interactive unbinding and rebinding of amprenavir into HIV-1 protease. The image in the upper left shows the bound "start" pose. (a) – (c) shown snapshots as the user unbinds amprenavir; (d) shows the unbound pose; and (e) – (g) shows snapshots as the user rebinds amprenavir. The corresponding RMSD time traces are shown in the middle plot, identical to that shown in Fig 3c.

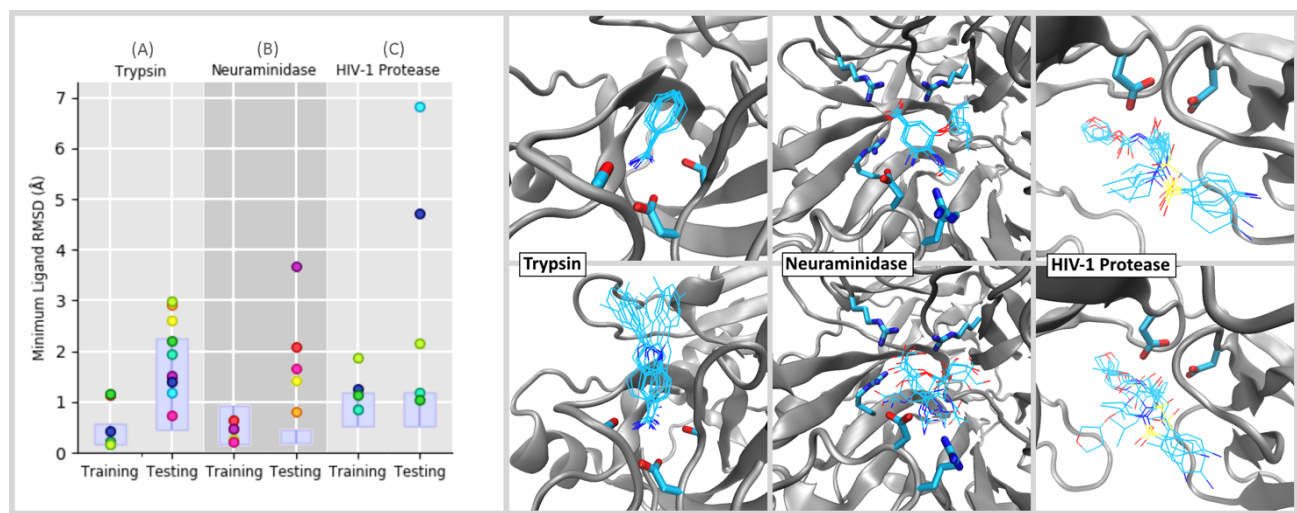

Figure 6: **Minimum RMSD values for novice tasks.** (Left) Showing the minimum achieved RMSD of each small ligand novices tried to dock, where each participant is assigned a unique colour. Two ligands were docked for each system, in both 'training' and 'testing' phases; each ligand is denoted in Figure 2. To determine the extent to which user-identified poses aligned with the native RMSD, a non-interactive Narupa simulation was run for 50 picoseconds in the bound pose and the average RMSD (plus or minus two standard deviations) was calculated, shown as the purple box. The whiskers at the top of each box span 1 Angstrom. (Right) The pose corresponding to the minimum RMSD value shown is on the left for each participant, overlaid on top of one another. Testing tasks for the three are shown in the top row, training tasks are shown on the bottom.

**3.2 Novice unbinding and rebinding tasks**

Figure 6 shows the minimum achieved RMSD of each small ligand novices tried to dock in both 'training' and 'testing' phases. For fully flexible simulations like these, which explore a range of conformational space as a result of thermal fluctuations, there is an open question how to assess whether a particular simulation has successfully recovered the binding pose seen in the crystal structure. For example, in a recent review by Pagadala *et al.* [57] which evaluated several different docking software programs, they used a cutoff of 2 Å to determine whether or not a fully flexible docking program managed to find the 'correct' pose. This strategy makes sense for cases where the protein's unbiased fluctuations span a range less than 2 Å. The solid boxes in the left hand panel of Figure 6 show the RMSD range (mean ±2σ) spanned by a non-interactive Narupa simulation run for 50 picoseconds initialized in the bound pose at 298 K. Fig 6 shows that the ±2σ range is less than 2 Å for all but the trypsin + indole-aminidine testing system. During user 'training' phases, the results from Figure 6 show that all of the user generated poses are either within ±2σ of the mean RMSD, or else have an RMSD which is within 2 Å of the crystal structure. During user 'testing' phases, where the binding guide was switched off and the ligand identity was different for trypsin and neuraminidase, the results show more scatter, exactly as we would expect. Nevertheless the results are encouraging. For trypsin, all of the novice-generated testing poses lie within ±2σ of the mean RMSD. For neuraminidase and HIV-1 protease, which are considerably more complicated tasks, the results are similarly encouraging: 4/5 of the neuraminidase testing poses are within 2.08 Å RMSD of the crystal structure; and 3/5 of the HIV-1 protease testing poses are within 2.15 Å RMSD of the crystal structure. Below we discuss the results for each system in further detail.

*3.2.1 Trypsin*

Directly comparing the trypsin training and testing tasks for the same participant shows that task accomplishment was better for all participants (ranging between 1.0Å and 1.9Å lower) during the training task. Benzamidine, used during the training task, is a small ligand and users had the benefit of trace atoms (Figure 6), explaining the good performance on the task. During the testing task, Figure 6 shows that participants were all able to recreate the amidine contact to Asp189, correctly aligning the indole-amidine benzamidine moiety. The variation in RMSD can be attributed to the rest of the molecule, where users had no prior guidance; variation in the extended scaffold between users can be seen in Figure 2A. Therefore, task accomplishment for binding indole-amidine into trypsin could likely be improved by providing additional feedback to indicate poses that are energetically favourable.

*3.2.2 Neuraminidase*

Figure 6 shows the minimum achieved RMSD for the neuraminidase training and testing tasks. All participants were able to recreate the original bound pose of oseltamivir in neuraminidase, getting within 1Å RMSD of the starting coordinates. The training results gave better minimum ligand RMSDs compared to the testing task, where were between 0.3Å and 3.2Å higher when the results between the training and testing task are compared for the same participant (Figure 6). For the testing task, four out of five participants were able to place zanamivir in the correct orientation, indicating that users recognized the shared scaffold between oseltamivir and zanamavir (highlighted in green in Figure 2B). In particular, there was very good alignment of polar moieties (e.g. the carboxylate group, the ring ether) between participants (Figure 6). However, variation arose from the flexible nature of zanamivir (whereas oseltamivir was very consistently aligned throughout the molecule when participants completed the task with the aid of trace atoms). This would suggest that an interface for completing drug binding tasks using iMD-VR could benefit from being able to give ligand-centric feedback, indicating whether functional groups are in an optimal geometry.

*3.2.3 HIV-1 Protease*

Figure 6C shows the minimum achieved RMSD for the HIV-1 protease training and testing tasks. Likely owing to the increased size of amprenavir and the number of rotatable bonds it has, both the training and test tasks for HIV-1 protease had a higher degree of variation in minimum RMSD achieved when directly comparing results between the two tasks for the same participant. Interestingly however, when rebinding amprenavir a second time without trace atoms to guide them (testing task), three participants were able to get an RMSD close to the training task, one of which was lower by 0.08Å. This may be a result of users remembering the correct binding pose but it is nonetheless encouraging that iMD-VR can be used to recreate poses of larger, torsionally complex molecules with limited iMD-VR training. In particular, recreation of the contact between the amprenavir hydroxyl group and Asp25B was generally good between both tasks (Figure S4-C). For the training task, Baker-Hubbard hydrogen bond analysis [58] on the minimum RMSD pose confirms this all participants were able to establish this contact. In the testing task, 4/5 participants were able to orientate the hydroxyl group and subsequently form a hydrogen bond to the catalytic Asp25A/Asp25B catalytic residues. However, the HIV-1 protease 'testing' task still had the highest variation in minimum RMSD, as some participants had a poorer performance without having a trace of the correct binding pose in the active site (the highest difference in RMSD between training and testing tasks was 5.9Å). In the testing task, four participants were able to orientate the hydroxyl group and subsequently form a hydrogen bond to the catalytic Asp25A/Asp25B catalytic residues. However, when inspecting the two outliers, the overall scaffold of amprenavir is incorrectly orientated – regardless of whether the user had established the hydrogen bonding contact to Asp25A and Asp25B.

## 4. Conclusions & Future Directions

In this paper, we have outlined an experimental protocol for setting up an iMD-VR simulation for the purposes of interactively manipulating protein-ligand systems to recover experimentally derived bound poses using iMD-VR. Utilizing this protocol, we have carried out studies exploring iMD-VR as a strategy for interactively sampling the unbinding and rebinding of ligands from proteins. The iMD-VR strategy enables this process to be accelerated compared to the much larger timescales required in unbiased MD simulations. To the best of our knowledge, this study represents the first time that iMD-VR has been extended to studies of complex protein-ligand binding and unbinding dynamics.

Our results show that expert iMD-VR users are able to manipulate protein-ligand systems to sample bound and unbound states. We also assessed the extent to which novice iMD-VR uses were able to sample binding and unbinding pathways using a training phase followed by a testing phase. Overall, the training phase showed slightly better task accomplishment; however, binding was still generally successful for many of the users during the testing phase, with the majority of users able to get within 2.15 Å RMSD of the experimental crystal structure, which is a comparable level of accuracy to other fully flexible docking programs. [57] So as not to overwhelm participants with having to learn a new VR-rendering interface during training, we took the decision to simplify the protein representations as shown in Supplementary videos A – C, showing a backbone representation along with a subset of those amino acids which are known to have key interactions with the ligands, and which make a large contribution to the overall binding free energy. Encouragingly, even though only a limited set of binding interactions were shown, participants were still able to accurately rebind the ligands studied herein, including amprenavir, whose size and flexibility makes the binding task particularly complicated. This raises an interesting question: Is it the quantity of rendered interactions (i.e., showing all possible interactions) or the quality of rendered interactions (i.e., showing a few of the most important interactions) which enable better results to be obtained during interactive docking tasks? At this stage, because we have not carried out the appropriate control experiment, our results do not conclusively show that focusing on a few key interactions enables better results, but this is an interesting avenue of future research which we intend to explore in future work. Nonetheless, these results are especially encouraging given that our cohort of novice users had a very limited time (less than 60 minutes) to be trained in both the interface and each of the three protein-ligand systems. The results suggest that the iMD-VR paradigm is sufficiently intuitive and affords adequate control to enable the sorts of detailed manipulations required to carry out VR-enabled atomistic docking in a fully flexible MD environment. Our 'expert' results show that iMD-VR users benefit from further training, given their familiarity with the interface and the specific protein systems.

Beyond the quantitative analysis described herein, these studies also show that iMD-VR can be used to reveal interesting qualitative features of protein-ligand unbinding, such as benzamidine not forming electrostatic contacts to Asp189 if it is incorrectly orientated and subsequently being repelled out of the trypsin binding pocket (Supplementary video A.2), or changes in the binding of amprenavir to tautomers of HIV-1 protease. In particular, the three-dimensional interface aided in the identification of key interactions between the protein and the ligand, as well as allowing the user to directly observe dynamical behaviours in real time.

Our iMD-VR framework enables users to quickly apply restraints to arbitrary subsets of atoms, which can be turned on and off 'on-the-fly'. The flexible application of restraints enables the user to rapidly identify those regions where they would like to either explore or dampen conformational flexibility – e.g., in loop domains. As a training strategy, we have found that restraints enable less experienced iMD-VR users to familiarize themselves with a system without the risk of inadvertently damaging the tertiary structure. For expert iMD-VR users, restraints permit the sampling of sophisticated docking strategies that involve complex conformational changes, like that in HIV-1 protease. In future work, we intend to carry out more detailed studies on how to apply restraints so as to enable users to efficiently and accurately carry out binding tasks.

We also intend to develop additional strategies to help iMD-VR users quickly formulate and test binding hypotheses. For example, checkpoints will enable users to return the system to previous states visited earlier in the simulation, along with additional audio and visual feedback to indicate to the user how structural manipulation influences the overall energy of the system, and allowing the user to pause the simulation and carry out energy minimizations 'on-the-fly'. For the purposes of streamlining these particular studies and not overwhelming novice users, we limited the studies described herein to a 'backbone and key residue' visualization scheme for the proteins. While such a representation aims to facilitate locating the active site and identifying key interactions, it makes it difficult to see unfavourable van der Waals interactions between the ligand and protein. In future work, we will examine the extent to which different rendering strategies influence task accomplishment. Such methods will help inform binding hypotheses in cases where the bound pose is unknown.

Unbiased simulation of protein ligand-binding typically occur on timescales of milliseconds or seconds [55, 56], and is therefore inaccessible to even the most sophisticated simulation architectures without recourse to a biasing or acceleration method. There is evidence that finding minimum energy pathways in hyperdimensional systems such as these is an "NP-hard" problem (for which no optimal method exists). [59] Having established herein that iMD-VR offers a reliable tool for experts and novices alike to generate accurate protein-ligand binding *poses*, we intend to carry out studies aimed at analysing the user-generated iMD-VR *pathways*. Specifically, we plan to analyse: (1) how closely user-generated iMD-VR pathways follow the system's minimum free energy path (MFEP); and (2) methods where iMD-VR pathways may be used to recover free energies for association and dissociation. For example, the configurations and pathways generated from iMD-VR (shown in Fig 3) could serve as initial guesses for adaptive path-based free energy sampling algorithms like transition path sampling [60], umbrella sampling [61], path-based metadynamics [62], forward flux sampling, [63] string methods, [64] or boxed molecular dynamics (BXD). [65] In addition, the configurations generated during an iMD-VR run could be used to produce Markov state models for ligand binding systems [8], by seeding the model with initial conditions which are not trapped within metastable energy minima [34, 66]. Whilst there are a number of details which require further investigation in order to establish a reliable and integrated workflow for recovering free energies along user-generated iMD-VR pathways, such a framework will enable us to explore the differences in kinetics and thermodynamics between different ligands binding to the same protein target, or the free energy landscape of the same drug binding to different protein mutants, with potential applications to areas like antimicrobial resistance. More broadly, the 3D iMD-VR interface can be used to explore and sample cryptic binding configurations which may not be indicated in the crystal structure, offering exciting new opportunities for interactive drug design.

**Competing Interests**

DRG acknowledges funding from: Oracle Corporation (University Partnership Cloud award). This does not alter our adherence to PLOS ONE policies on sharing data and materials.

# Supplementary Information

## 1. Data & Video files

The Open Science Framework (see the URL at http://doi.org/10.17605/OSF.IO/NCFQM) includes the raw simulation data utilized to generate all of the Figures in the main text and the SI, the associated Narupa input files, and the videos referred to within the text. The specific version of the Narupa software used to generate the simulations discussed in the text is available at https://irl.itch.io/narupaxr. The videos and animations discussed in the main text are also available on Vimeo at the following hyperlinks:

Supplementary Animation A: https://vimeo.com/354828618

Supplementary Animation B: https://vimeo.com/354829098

Supplementary Animation C: https://vimeo.com/354829412

Supplementary Video A.1: https://vimeo.com/354833443

Supplementary Video A.2: https://vimeo.com/380015176

Supplementary Video B: https://vimeo.com/354833800

Supplementary Video C: https://vimeo.com/354834013

## 2. System parameterization

Systems were selected where an existing crystal structure in PDB format exists. For trypsin and benzamidine, the bound complex coordinates were based on PDB file 1S0R. For HIV-1 protease and amprenavir, the bound complex coordinates were based on PDB file 1HPV. For neuraminidase and oseltamivir, the bound complex coordinates were based on the starting structures used in the production runs in previous work by Woods *et al.* (which were themselves based on PDB file 2QWK) [1]. For the user studies described in the main text, two additional ligand systems were identified: trypsin and indole-amidine, based on PDB file 2G5N; and neuraminidase and zanamivir, based on PDB file 6HCX.

For performing iMD-VR runs, all proteins were parameterized with the Amber 14ffSB force field[2]. Benzamidine, indole-amidine, zanamivir, and amprenavir were parameterized with GAFF using antechamber. If needed, hydrogens were added to the systems using reduce [3]. Parameters for oseltamivir were taken from Woods et al [1]. All three systems used the OBC2 implicit water model. All systems were energy minimized before being used as the starting coordinates for each bound complex. All iMD-VR simulations were run at 300K with a timestep of 0.5 fs. When interacting with atoms, we utilized a gaussian force, where the amount of force applied depends on the distance between the user's controller and the atom they were selecting. Once users stopped interacting with a selection, the velocities of all atoms in that selection were reinitialized according to a Boltzmann distribution based on the temperature in order to prevent any remaining interaction energy in molecules propelling them past where the user intends to place them. Further details of how the velocities are reset can be found in ref [4].

For the expert-level tests, an 800 kJ/mol/nm$^2$ restraint was added to all backbone atoms in both trypsin and neuraminidase; additionally, this restraint was applied to any Ca2+ cations embedded within both systems. An 800 kJ/mol/nm$^2$ was applied to all backbone atoms in HIV-1 protease, excluding those which make up the flaps which gate the active site (defined as residues 49 to 55 in chain A and residues 48 to 54 in chain B). A separate force was applied to the HIV-1 flap backbone atoms which, unlike the other backbone restraint, the user could turn on and off during the interactive simulation. The software default value of 2000 kJ/mol/nm2 force constant was used for these restraints. A higher force constant is used compared to the backbone restraints due to their interactive nature. When a user wishes atoms to be held in place, it is expected that they want to be held in place firmly. All training and testing non-expert user tasks for trypsin and hiv-1 protease had the same backbone restraints as the expert tests; these restraints were removed for the neuraminidase blind task for one two novice cohorts that were recruited. The other cohort used the same backbone restraints as the expert-level test. In future work, we intend to carry out a more detailed sensitivity analysis of different restraint regimes.

## 3. NarupaXR input files & settings

Input files for every system discussed in this work can be downloaded from the ESI. The input files consist of a PDB structure showing the starting bound coordinates, an .xml file which contains the simulation information to be loaded into NarupaXR, and a .json file

## 4. Equilibration protocols

For three of the protein systems, a single frame (the minimum RMSD along the iMD-VR trajectory) was extracted from the simulation trajectories for production run molecular dynamics, whose results are shown in Fig 4. The extracted frame from trypsin had an RMSD of 0.041 for benzamidine compared to the starting coordinates. For neuraminidase the extracted frame had an RMSD of 0.27 for oseltamivir and 0.302 for the 150-loop compared to the starting coordinates. For HIV-1 protease, the extracted frame had an RMSD of 0.218 for amprenavir and 1.383 for the active site flaps compared to the starting coordinates.

Extracted frames were solvated and neutralized: 10515 water molecules and 9 Cl- ions were added to trypsin and benzamidine, 17956 water molecules and 1 Cl- ion were added to neuraminidase and oseltamivir, 10696 water molecules and 5 Cl- ions were added to HIV-1 protease and amprenavir. All water was treated with the TIP3P water model. Forcefields for the protein and ligand were kept consistent between the interactive simulation and production run.

The solvated frames were minimized and equilibrated using the following protocol. Post minimization, the system was linearly heated over 20 ps, starting at 30K and increasing the temperature by 30K every 2 ps, until a temperature of 298K was reached. Next, a 10 kcal/mol/Å$^2$ restraint was added to the protein backbone atoms and the simulation was run for five sets of 10ps under the NVT ensemble, reinitializing the velocities of atoms in the system before each stage. Next, any restraints on the protein were removed and the box volume was allowed to relax to a pressure of 1 bar by running under the NPT ensemble for 9 nanoseconds. Finally, the system was switched back to the NVT ensemble, before being run for a final nanosecond. We note that there are a number of ways which we could have chosen in order to initiate the subsequent 200 ns MD simulations, and it is unclear *a priori* which choice is optimal. We chose the minimum simply because it is uniquely defined and unambiguous. Given that the snapshots which we extracted for production molecular dynamics were first solvated, minimized and then equilibrated for 10ns, the minimum RMSD structure would have been immediately relaxed to other nearby fluctuating structures; it is therefore unlikely that choosing the minimum as the starting point gives a markedly different result than choosing some other average structure nearby.

The final box dimensions are as follows: For trypsin and benzamidine the dimensions were 69.109 by 66.786 by 75.289, for neuraminidase and oseltamivir the dimensions were 88.788 by 81.528 by 82.755, and for HIV-1 protease and amprenavir the final box dimensions were 68.894 by 65.427 by 78.456.

## 5. RMSD calculation

Narupa requires a set of atom coordinates from which to begin an iMD-VR simulation. These starting coordinates were used as the reference frame for the RMSD calculation. As such, the RMSD was calculated as a measure of the distance the ligand has moved compared to the starting frame (i.e. when it is in a bound position). To calculate the RMSD for the ligand, the protein backbone atoms in each step of the trajectory were aligned to those in the starting coordinates. Once the trajectory was aligned, the RMSD was taken for the ligand atoms compared to the starting coordinates (excluding hydrogens). However, an additional consideration is that all molecules tested display symmetry in some of their functional groups. Narupa allows users to quickly explore rotational orientations, which means that users can flip the orientation of certain rotationally invariant groups, and inadvertently introduce a degree of variation into RMSD calculations. Therefore, when calculating the RMSD of the ligand, symmetrical groups were identified (see figure S1) and the swap plugin of VMD was used to calculate a more optimal RMSD value.

Additionally, to give an idea of the RMSD when docked, a 'baseline' RMSD was calculated by allowing each system to run for 50 ps in Narupa without any perturbation from the user, using a timestep of 0.5 fs and recording interval of 0.25 ps, totalling 200 RMSD data points. In order to gain an idea of the native behaviour of each ligand within a fully flexible protein, the backbone restraints were relaxed to 5 kJ/mol/nm$^2$. In total, 60 ps of simulation time was recorded, however, the first 10 ps was discarded to give the simulation the opportunity to relax itself. Any ions in the system used backbone restraints of 800 kJ/mol/nm$^2$. As with above, the RMSD was taken excluding the position of any hydrogens in the ligand. The RMSD value was either taken as (a) a single mean value over this 'reference' trajectory or (b) a range, calculated as plus or minus two standard deviations from the mean. These values can be found in Table S2. A dataset containing the RMSD values for each reference trajectory of each system can be found in the supplementary files of the ESI.

## 6. NarupaXR technical specifications

Narupa consists of two components: A molecular dynamics server engine that propagates a simulation forward and transmits the atomistic data to a client application, which renders the simulation in VR, and provides the user interface. For small to moderate sized protein simulations, these two components can be run locally on the same machine. Alternatively, a server machine can run the simulation engine and transmit the atomistic data to other VR clients, allowing multiple people to interact with the same iMD-VR simulation. For the expert-level user tests, both the molecular dynamics engine and VR front end were run on a high-end Alienware 15 R4 gaming laptop with an Intel i7-8750H 2.20GHz CPU and a NVIDIA GeForce GTX 1070 graphics card. For non-expert user tests, both the molecular dynamics engine and VR front end were run on a high-end desktop PC with an AMD Ryzen 31200 3.10GHz CPU and a NVIDIA GeForce GTX 1060 GPU. During the novice user tests, an expert explained the protein system to the user in a multiplayer simulation, which required two **client** machines, with the data from the simulation engine transmitted to a computer with identical specifications over a local area network.

### 7. Participant-generated bound poses

Figure 6 in the main text shows an overlay of the minimum RMSD bound pose achieved by each of the five novice participants, showing the increased variation when participants had ghost atoms (top three images) and when they did not (bottom three images). Additionally, for the final 5 ps of the participant-generated trajectories in cohort 2, the distance between atoms in the ligand molecule with key atoms in the protein residue are shown in Figure S4. Table S1 shows the minimum RMSD of each participant for each task.

### 8. Generation of HIV-1 open flap coordinates

In order to successfully recreate the opening and closing process of HIV-1 protease, an indication of the correct open flap position was created by creating a ghost atom representation of the backbone atoms of open HIV-1 protease (PDB code: 1TW7) and then superimposing that structure over the HIV-1 protease starting coordinates. Next, an iMD-VR user took advantage of the three-dimensional capabilities of Narupa and manually inspected the 'hinge-point' of the two flaps, where the positions of backbone atoms in the open and closed positions start to diverge. From this, a trace atom representation was created with two sets of ghost atoms, one for the closed and one for the open position, giving users a sense for the normal range of movement of the flaps. The areas of the protein backbone encapsulated by the hinge points were defined as between residues 48 and 54 on both protein chains.

### 9. Docking and redocking without backbone restraints

After establishing the usability of our framework, the experts repeated the tests with the 800 kJ/mol/nm$^2$ backbone restraints removed from all simulation files. Experts were able to manipulate all three systems with minimal interference to the protein tertiary structure, indicating that backbone restraints can potentially be removed. However, an added benefit of using relatively high backbone restraints is that paths generated are more straightforward to interpret because the motion from thermal fluctuations is curtailed. As a result, PathReducer provides principle components which better capture the key features of user-sampled iMD-VR pathways (i.e. the movement of a ligand from a protein binding site rather than side chain fluctuations). For cases where loop displacement is important (e.g., the motion of residues in the 150-loop during the novice neuraminidase testing tasks), this can be a potential contributor to poor task accomplishment. For example, comparing the re-establishment of the 150-loop for the neuraminidase testing task where both backbone restraints are on and off, figure S5 shows that the RMSD of these key residues is poorer and more varied when backbone restraints are not present. Furthermore, Figure S4 shows that contact between the zanamavir carboxylate group and Arg371 (one of the trio of arginines which is important for binding) is consistently poor, indicating non-optimal placement of the ligand by participants - which is further supported by figure S3, which shows placement of the carboxylate group as better where backbone restraints are present. This would suggest that where placement of key residues is important, such as the 150-loop, backbone restraints have the benefit of restricting the degrees of freedom in a protein. For novice users who are unfamiliar with both iMD-VR controls and the protein system itself, a high backbone restraint gives users the ability to familiarize themselves with the process of undocking and redocking without fear of accidentally pushing the protein system into a state where binding is unfavorable.

### 10. PathReducer analysis

*PathReducer* takes as input an xyz file containing a series of molecular structures (in this case, an iMD-VR-generated trajectory) and uses principle component analysis to determine the optimal rotation and scaling of the original coordinate system that captures the most structural variance in the fewest coordinates. This simple rotation and scaling means that the resultant coordinate system is comprised of coordinates that are linear combinations of the input coordinates (or input 'features'). By looking at the coefficient (or 'weight') of each input coordinate in the new coordinate system, one can see the relative amount that input coordinate is contributing to the direction of the new coordinate. The projections of the original data onto the resultant coordinates are referred to as principal components (PCs). Plots of iMD-VR pathways in PC space provides an idea of the similarity of structures sampled along the pathway. These plots can also be used to understand the 'route' taken by a trajectory. Figure 3 of the main text shows the iMD-VR-generated binding and unbinding pathways in the space defined by their top two and top three PCs. The structures input into *PathReducer* were represented as squared interatomic distance matrices, owing to their rotational and translational invariance compared to Cartesian coordinates.

For computational tractability, the PCA considered only those 70,000 interatomic distances that varied the most along a trajectory (only taking into account non-hydrogen atoms). In the plots in Figure 3 of the main text, purple corresponds to the beginning of the trajectory and yellow corresponds to the end, with blue and green passed through in between. Each scatter point is a frame from the iMD-VR trajectory, and a linear interpolation was used to connect each point.

## 11. Demographic Information

Novice users were recruited in two cohorts of five, totalling ten participants total. Cohort one completed the testing and training tasks for trypsin and neuraminidase, where all protein systems had a restraint placed on their backbone atoms. Cohort two completed the testing and training tasks for trypsin, neuraminidase and HIV-1 protease, however, for the neuraminidase testing task, all restraints were turned off. For cohort one, of the five participants, three participants were female and one was male. One participant opted not to specify their gender. Four participants were between the ages of 18 and 24 and one was between the age of 25 and 34; one participant was an undergraduate taught student, one participant was a postgraduate taught student, and three were postgraduate research students. Prior experience with VR was low amongst the cohort: three participants rated themselves as having a little experience using VR and one participant rated themselves as somewhat experienced. One member of the cohort rated themselves as very experienced. For cohort two, of the five participants, four were male and one was female; two participants were between the ages of 18 and 24 and three were between the ages of 25 and 34; two participants were postdoctoral researchers, two were postgraduate research students and one was a postgraduate taught student. Prior experience with VR was low amongst the cohort: two participants rated themselves as having no experience using virtual reality, two participants rated themselves as being slightly experienced, and one participant rated themselves as somewhat experienced.

## 12. Additional figures and tables

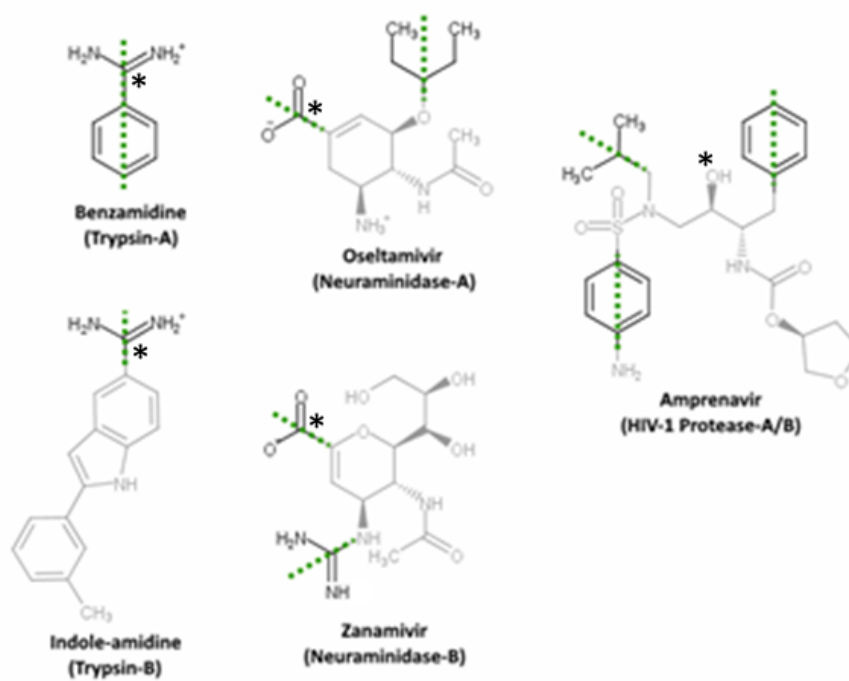

**Figure S1** The symmetrical groups in each of the interactively undocked and redocked molecules. In these cases, the swap plugin of VMD was used so they could be treated as rotationally invariant when calculating RMSD. The atoms labelled with * are the key ligand atoms which participate in the distances shown in Fig S4.

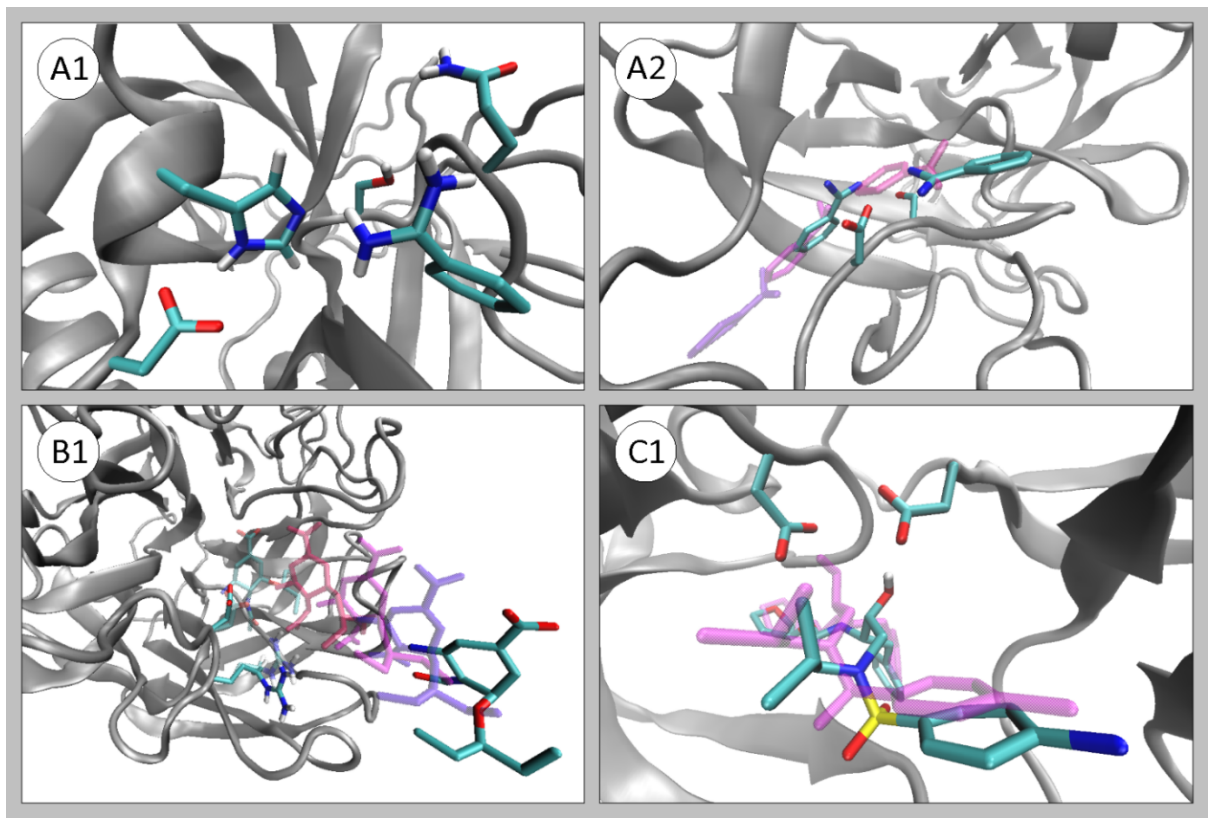

**Figure S2** Snapshots of iMD-VR generated trajectories that illustrate observations by the expert level users. (A1) Shows the binding pose of incorrectly orientated benzamidine after being repelled from the trypsin binding pocket, in particular, how it binds to the catalytic Ser-His-Asp triad of benzamidine (plus neighbouring Glu192) on the way out, as shown in Supplementary video A.2. (A2) shows the path trypsin takes when being moved by an expert iMD-VR user through the hydrophobic protein core. In this case, starting from the docked pose, the user was able to pull benzamidine into a binding mode where it instead sits underneath Asp189 from this pose; the user was then able to continue pulling benzamidine until it exited the protein through an alternative pocket. Snapshots demonstrating the motion of the ligand are rendered as semi-transparent purple (B1) shows oseltamivir flipping its orientation when being pulled out of the neuraminidase binding pocket. Snapshots demonstrating the motion of the ligand oseltamivir as rendered as transparent purple; additionally, the initial positions of oseltamivir and the 150-loop are rendered as semi-transparent and in the CPK colouring style. (C1) shows an alternate binding pose of amprenavir into a tautomer of HIV-1 protease, where neither of the catalytic aspartic acid residues are protonated, demonstrating how the structure of amprenavir sags when the hydroxyl group switches which aspartic acid it points towards. The starting binding pose is rendered as semi-transparent and purple.

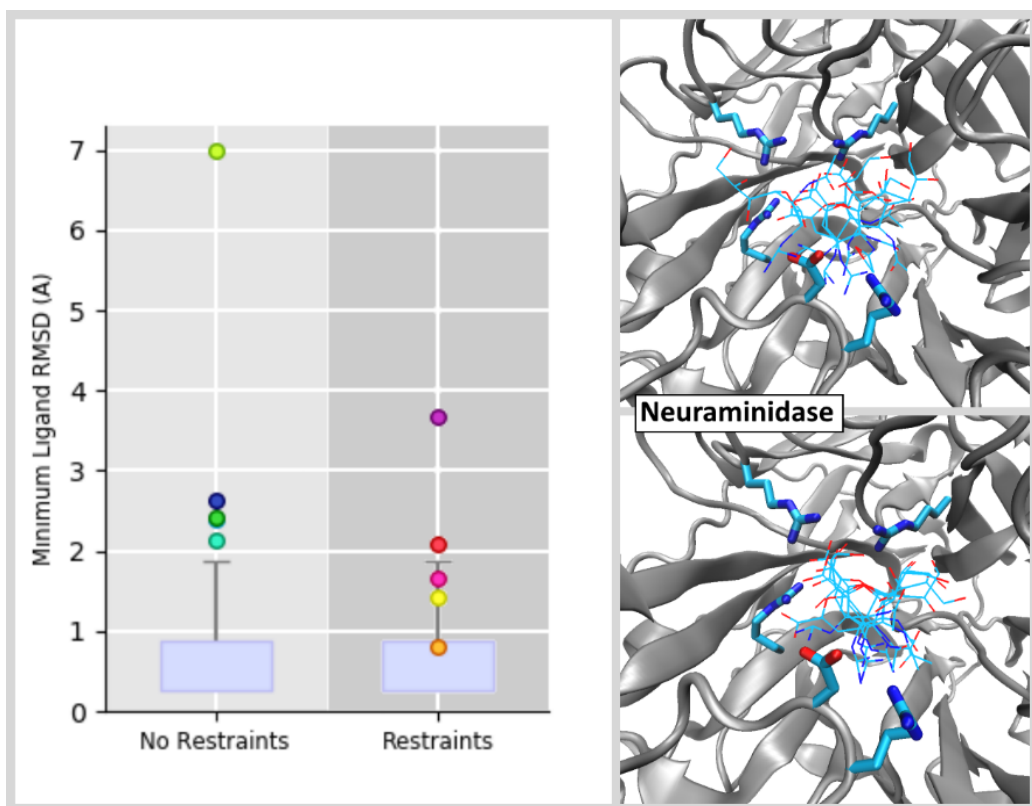

**Figure S3** (Left) Minimum achieved RMSD for zanamivir in neuraminidase (testing task), where five participants completed the task without backbone restraints (cohort one) and with backbone restraints (cohort two). Each participant is assigned their own colour). (Right) Bound poses corresponding to minimum RMSDs shown on the left. No restraints is the top panel, restraints is the bottom panel.

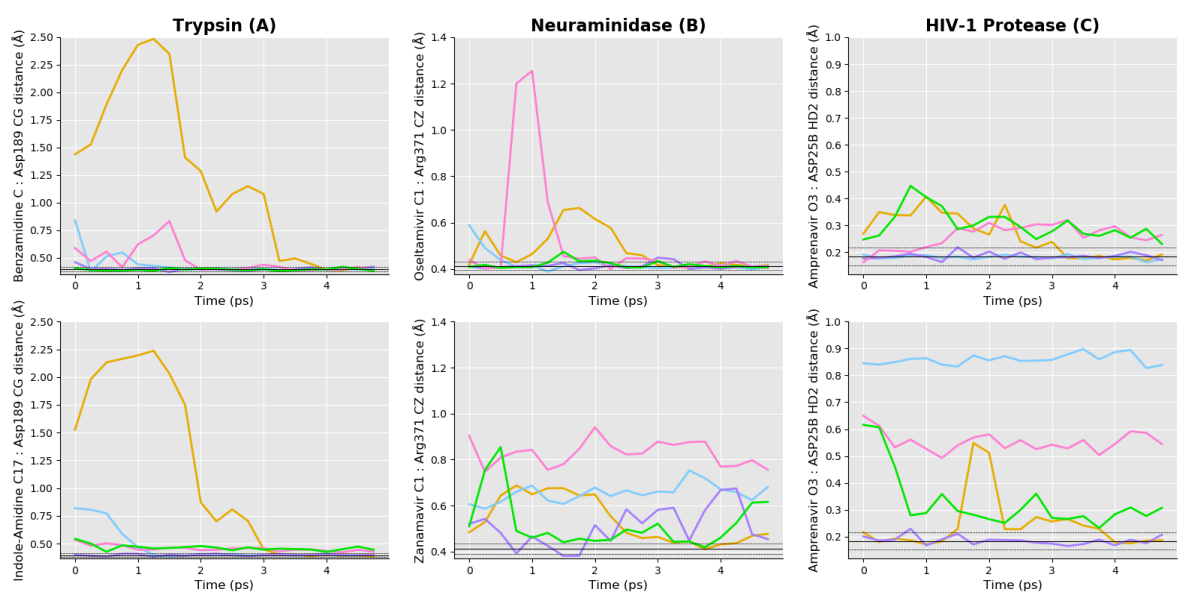

**Figure S4**: atom distances between key ligand atoms (labelled with * in Fig S1) and key protein residues (shown in figure 1 and 2 of the main text) through the final 5 ps of trajectories generated by 5 user study participants from cohort two, where each color represents the results generated by a different participant. The top row shows the results for training tasks, and the bottom row shows the results for testing tasks. The solid black line shows the system-specific average (with black dotted lines representing ±2σ) for a non-interactive Narupa simulation run for 50 ps in the bound pose and the calculated average atom distance is shown by the blue line. For benzamidine and indole-amidine, the distance represents the contact between their amidine group and Asp189 of trypsin. For oseltamivir and zanamavir, the distance represents the contact between their carboxylic acid group and Arg371. For amprenavir, the distance represents the contact between the hydroxyl oxygen and the protonated catalytic residue of HIV-1 protease, Asp25B.

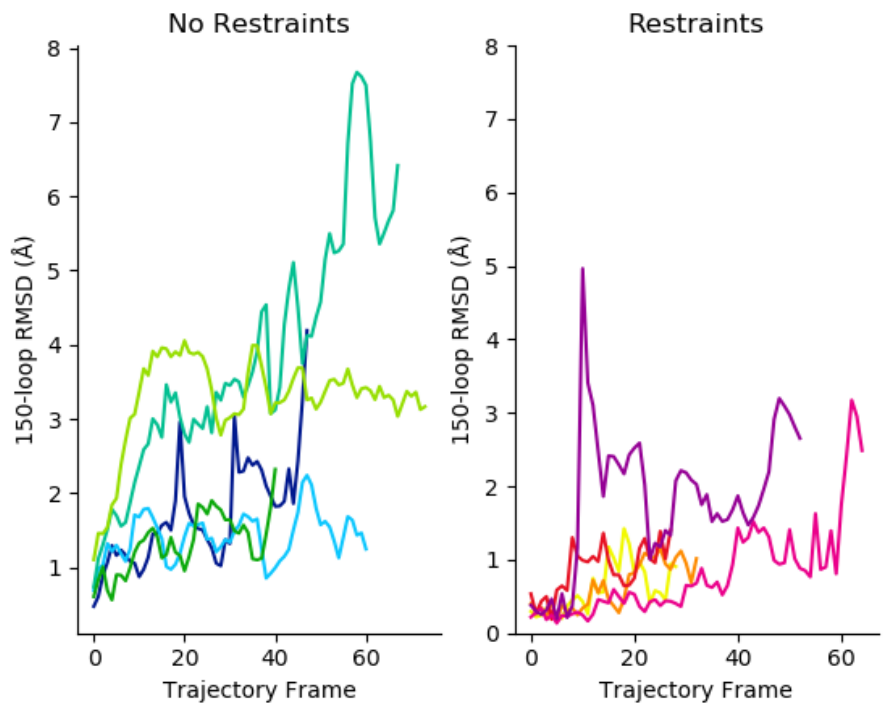

**Figure S5** shows the RMSD of the two 150-loop residues, Asp151 and Arg152, of neuraminidase for the testing task, both with and without backbone restraints, showing that the RMSD of the 150-loop without restraints was (a) higher and (b) exhibited a greater degree of variation compared to with restraints. Each participant is assigned their own color.

| Participant | Ligand RMSD (Å) | | | | | | |
| --- | --- | --- | --- | --- | --- | --- | --- |
| | A training | A testing | B training | B testing (restraints) | B testing (no restraints) | C training | C testing |
| C1 – 1 | 0.23 | 2.61 | 0.27 | 1.41 | n/a | n/a | n/a |
| C1 – 2 | 0.18 | 2.91 | 0.48 | 0.80 | n/a | n/a | n/a |
| C1 – 3 | 1.13 | 1.49 | 0.64 | 2.08 | n/a | n/a | n/a |
| C1 – 4 | 0.20 | 0.74 | 0.21 | 1.67 | n/a | n/a | n/a |
| C1 – 5 | 0.23 | 1.53 | 0.48 | 3.68 | n/a | n/a | n/a |
| C2 – 1 | 0.43 | 1.39 | 0.58 | n/a | 2.63 | 1.26 | 4.71 |
| C2 – 2 | 0.16 | 1.20 | 0.73 | n/a | 2.40 | 0.85 | 6.82 |
| C2 – 3 | 0.20 | 1.93 | 0.81 | n/a | 2.13 | 0.85 | 1.19 |
| C2 – 4 | 1.15 | 2.20 | 0.74 | n/a | 2.42 | 1.14 | 1.04 |
| C2 – 5 | 0.18 | 3.00 | 0.87 | n/a | 6.99 | 1.88 | 2.15 |

**Table S1** The reported minimum ligand RMSD values for each participant for each task. Protein systems are categorized by letter, where A is trypsin, B is neuraminidase and C is HIV-1 protease. Participants are sorted by cohort, where C1 denotes the first cohort and C2 denotes the second cohort. Where a participant did not complete a given task, the RMSD value is n/a.

| Participant | Ligand RMSD (Å) | | | | | | |
| --- | --- | --- | --- | --- | --- | --- | --- |
| | A training | A testing | B training | B testing (restraints) | B testing (no restraints) | C training | C testing |
| C1 – 1 | 0.34 | 2.81 | 0.27 | 1.44 | n/a | n/a | n/a |
| C1 – 2 | 0.38 | 3.29 | 1.22 | 1.12 | n/a | n/a | n/a |
| C1 – 3 | 1.16 | 2.70 | 0.81 | 3.47 | n/a | n/a | n/a |
| C1 – 4 | 0.50 | 1.02 | 0.52 | 2.47 | n/a | n/a | n/a |
| C1 – 5 | 0.39 | 3.23 | 0.57 | 6.34 | n/a | n/a | n/a |
| C2 – 1 | 0.47 | 2.35 | 0.58 | n/a | 3.22 | 1.86 | 5.20 |
| C2 – 2 | 0.29 | 1.81 | 0.93 | n/a | 2.81 | 1.25 | 7.46 |
| C2 – 3 | 0.62 | 1.93 | 0.91 | n/a | 7.59 | 1.23 | 1.24 |
| C2 – 4 | 1.19 | 3.40 | 0.74 | n/a | 4.00 | 2.54 | 1.15 |
| C2 – 5 | 0.24 | 3.33 | 1.20 | n/a | 8.92 | 2.16 | 2.55 |

**Table S2** The reported final ligand RMSD values for each participant for each task. Protein systems are categorized by letter, where A is trypsin, B is neuraminidase and C is HIV-1 protease. Participants are sorted by cohort, where C1 denotes the first cohort and C2 denotes the second cohort. Where a participant did not complete a given task, the RMSD value is n/a.

| System | Average RMSD (± 2σ) (Å) |
|---|---|
| Benzamidine (A training) | 0.60 (± 0.30) |
| Indole-Amidine (A testing) | 2.14 (± 1.23) |
| Oseltamivir (B training) | 0.93 (± 0.47) |
| Zanamavir (B testing) | 0.56 (± 0.31) |
| Amprenavir (C testing and training) | 0.77 (± 0.40) |

**Table S3** The average RMSD of each ligand used in IMD-VR docking tasks, calculated over 50 ps of simulation time. Protein systems are categorized by letter, where A is trypsin, B is neuraminidase and C is HIV-1 protease. Two times the standard deviation is shown in brackets. Each value is given to three significant figures.

## Supplementary Information References